\documentclass[onecollarge,natbib]{svjour2}
\bibpunct{[}{]}{;}{n}{}{,} 
\smartqed  
\usepackage{graphicx}
\usepackage{amsmath}
\usepackage{amsfonts}
\usepackage{amssymb}
\newcommand{\bbox}[1]{\mbox{\boldmath $#1$}}
\newcommand{\boldtau}{\mbox{\boldmath $\tau$}}
\newcommand{\beq}{\begin{equation}}
\newcommand{\eeq}{\end{equation}}
\newcommand{\beqa}{\begin{eqnarray}}
\newcommand{\eeqa}{\end{eqnarray}}
\newcommand{\eq}[1]{Eq.~(\ref{#1})}
\journalname{Few-Body Systems}
\begin{document}

\title{Infinite-cutoff renormalization of the chiral nucleon-nucleon interaction up to N$^3$LO
}
\dedication{Dedicated to Professor Henryk Witala on the occasion of his 60th birthday.}


\author{Ch. Zeoli   \and  R. Machleidt  \and  D. R. Entem
}


\institute{Ch. Zeoli \at
              Department of Physics, University of Idaho, Moscow, Idaho 83844-0903, USA \\
                          \emph{Present address:} Department of Physics, Florida State University,
                          Tallahassee, Florida 32306-4350, USA \\
                           \email{cpz11@my.fsu.edu}  
           \and
           R. Machleidt \at
            Department of Physics, University of Idaho, Moscow, Idaho 83844-0903, USA \\
               \email{machleid@uidaho.edu}    \\
               \and
               D. R. Entem \at
               Grupo de Fisica Nuclear and IUFFyM, University of Salamanca, E-37008 Salamanca, Spain \\
                \email{entem@usal.es}  
}

\date{Received: date / Accepted: date}

\maketitle

\begin{abstract}
Naively, the ``best'' method of renormalization is the one where a momentum cutoff is taken to infinity while maintaining stable results due to a cutoff-dependent adjustment of counterterms. We have applied this renormalization method in the non-perturbative calculation of phase-shifts for nucleon-nucleon ($NN$) scattering using chiral $NN$ potentials up to next-to-next-to-next-to-leading order (N$^3$LO). For lower partial waves, we find that there is either no convergence with increasing order or, if convergence occurs, the results do not always converge to the empirical values. For higher partial waves, we always observe convergence to the empirical phase shifts (except for the $^3$G$_5$ state). Furthermore, no matter what the order is, one can use only one or no counterterm per partial wave, creating a rather erratic scheme of power counting that does not allow for a systematic order-by-order improvement of the predictions. The conclusion is that infinite-cutoff renormalization is inappropriate for chiral $NN$ interactions, which should not come as a surprise, since the chiral effective field theory, these interactions are based upon, is designed for momenta below the chiral-symmetry breaking scale of about 1 GeV. Therefore, this value for the hard scale should also be perceived as the
appropriate upper limit for the momentum cutoff.
\end{abstract}

\section{Introduction}
\label{sec_intro}

During the past two decades, it has been demonstrated that chiral effective field theory (chiral EFT) represents 
a powerful tool to deal with hadronic interactions at low energy in a systematic and model-independent
way (see Refs.~\cite{ME11,EHM09} for recent reviews). 
For the construction of an EFT, it is crucial to identify a separation of
scales. In the hadron spectrum, a large gap between the masses of
the pions and the masses of the vector mesons, like $\rho(770)$ and $\omega(782)$,
can clearly be identified. Thus, it is natural to assume that the pion mass sets the soft scale, 
$Q \sim m_\pi$,
and the rho mass the hard scale, $\Lambda_\chi \sim m_\rho \sim 1$ GeV,
also known as the chiral-symmetry breaking scale.
This is suggestive of considering a low-energy expansion arranged in terms of the soft scale over the hard scale,
$(Q/\Lambda_\chi)^\nu$,  where $Q$ is generic for an external
momentum (nucleon three-momentum or pion four-momentum) or a pion mass.
The appropriate degrees of freedom are, obviously,  pions and nucleons, and not quarks and gluons.
To make sure that this EFT is not just another phenomenology,
it must have a firm link with QCD.
The link is established by having the EFT observe
all relevant symmetries of the underlying theory, in particular,
the broken chiral symmetry of low-energy QCD~\cite{Wei79}. 

The early applications of chiral perturbation theory (ChPT) focused on systems like $\pi\pi$~\cite{GL84}
and $\pi N$~\cite{GSS88}, where the Goldstone-boson character of the pion
guarantees that a perturbative expansion exists.
But the past 15 years have also seen great progress in applying ChPT to nuclear forces
\cite{ME11,EHM09,Wei90,Wei91,Wei92,ORK94,ORK96,Kol94,Kol99,KBW97,KGW98,EGM98,EGM00,BK02,EM02,EM02a,EM03,ME05,EGM05,ME10}. 
The nucleon-nucleon ($NN$) system is characterized by large scattering lengths and bound states indicating the non-perturbative
character of the problem. Weinberg~\cite{Wei90,Wei91} therefore suggested to calculate the nuclear amplitude
in two steps. In step one, the {\it nuclear potential}, $\widehat{V}$, is defined as the sum of irreducible
diagrams, which are evaluated {\it perturbatively} up to the given order. Then in step two, this potential
is iterated to all order (i.e., summed up {\it non-perturbatively}) in the Schr\"odinger or 
Lippmann-Schwinger (LS) equation:
\begin{equation}
 \widehat{T}({\vec p}~',{\vec p})= \widehat{V}({\vec p}~',{\vec p})+
\int d^3p''\:
\widehat{V}({\vec p}~',{\vec p}~'')\:
\frac{M_N}
{{ p}^{2}-{p''}^{2}+i\epsilon}\:
\widehat{T}({\vec p}~'',{\vec p}) \, ,
\label{eq_LS}
\end{equation}
where $\widehat{T}$ denotes the $NN$ T-matrix and $M_N$ the nucleon mass.

In general, the integral in
the LS equation is divergent and needs to be regularized.
One way to do this is  by
multiplying $\widehat{V}$
with a regulator function
\begin{equation}
\widehat{V}(\vec{ p}~',{\vec p}) 
\longmapsto
\widehat{V}(\vec{ p}~',{\vec p})
\;\mbox{\boldmath $e$}^{-(p'/\Lambda)^{2n}}
\;\mbox{\boldmath $e$}^{-(p/\Lambda)^{2n}}
\label{eq_regulator} \,.
\end{equation}
Typical choices for the cutoff parameter $\Lambda$ that
appears in the regulator are 
$\Lambda \approx 0.5 \mbox{ GeV} \ll \Lambda_\chi \approx 1$ GeV.

It is pretty obvious that results for the T-matrix may
depend sensitively on the regulator and its cutoff parameter.
This is acceptable if one wishes to build models.
For example, the meson models of the past~\cite{Mac89,MHE87}
always depended sensitively on the choices for the
cutoff parameters which, in fact,
were important for the fit of the $NN$ data.
However, the EFT approach wishes to be fundamental
in nature and not just another model.

In field theories, divergent integrals are not uncommon and methods have
been developed for how to deal with them.
One regulates the integrals and then removes the dependence
on the regularization parameters (scales, cutoffs)
by renormalization. In the end, the theory and its
predictions do not depend on cutoffs
or renormalization scales.
So-called renormalizable quantum field theories, like QED,
have essentially one set of prescriptions 
that takes care of renormalization through all orders. 
In contrast, 
EFTs are renormalized order by order. 

The renormalization of {\it perturbative}
EFT calculations is not a problem. {\it The problem
is non-perturbative renormalization.}
This problem typically occurs in {\it nuclear} EFT because
nuclear physics is characterized by bound states which
are non-perturbative in nature.
EFT power counting may be different for non-perturbative processes as
compared to perturbative ones. Such difference may be caused by the infrared
enhancement of the reducible diagrams generated in the LS equation.

Weinberg's implicit assumption~\cite{Wei90,Wei91,Wei09} was that the counterterms
introduced to renormalize the perturbatively calculated
potential, based upon naive dimensional analysis (``Weinberg counting''),
are also sufficient to renormalize the non-perturbative
resummation of the potential in the LS equation.
In 1996, Kaplan, Savage, and Wise (KSW)~\cite{KSW96,KSW98,KSW98a}
pointed out that there are problems with the Weinberg scheme
if the LS equation is renormalized 
by minimally-subtracted dimensional regularization.
This criticism resulted in a flurry of publications on
the renormalization of the non-perturbative
$NN$ problem
\cite{FMS00,FMS00a,PBC98,FTT99,Bea02,VA05-1,NTK05,VA05-2,Bir06,VA07,EM06,VA08,Ent08,YEP07,YEP08,YEP09,LK08,BKV08,Val08,Mac09,Val09,Val11,Tim11,LY12,Bir11}.
The literature is too comprehensive
to discuss all contributions.
Let us just mention some of the work that has particular relevance
for our present discussion.

If the potential $\widehat{V}$ consists of contact terms only (a.k.a.\
pion-less theory), then
the non-perturbative summation Eq.~(\ref{eq_LS})
can be performed analytically and the power counting is explicit.
However, when pion exchange is included, then Eq.~(\ref{eq_LS})
can be solved only numerically and the power counting
is less transparent.
Perturbative ladder diagrams of arbitrarily high order,
where the rungs of the ladder represent a potential made up from
irreducible pion exchange,
suggest that an infinite number of counterterms is needed to achieve
cutoff independence for all the terms of increasing order generated
by the iterations.
For that reason, 
KSW~\cite{KSW96,KSW98,KSW98a} proposed 
to sum the leading-order contact interaction to all orders (analytically)
and to add higher-order contacts and
pion exchange perturbatively up to the given order. Unfortunately,
it turned out that the order by order convergence of one-pion exchange (1PE) 
is poor in the $^3S_1$-$^3D_1$ state~\cite{FMS00,FMS00a}. 
The failure was triggered by the $1/r^3$ singularity of the 1PE tensor
force when iterated to second order. Therefore, KSW counting is no
longer taken into consideration (see, however, \cite{BKV08}).  
A balanced discussion of possible
solutions can be found in \cite{Bea02}.

Some researchers decided to take
a second look at Weinberg's original proposal.
A systematic investigation of Weinberg counting in leading order (LO)
has been conducted by Nogga, Timmermans, and van Kolck~\cite{NTK05}
in momentum space, and by Valderrama and Arriola
at LO and higher orders in
configuration space~\cite{VA05-1,VA05-2,VA07}. A comprehensive
discussion of both approaches and their equivalence can be found
in~\cite{Ent08,Val08}.

The LO $NN$ potential consists
of 1PE plus two nonderivative contact terms that contribute
only in $S$ waves.
Nogga {\it et al} find that the given counterterms renormalize
the $S$ waves (i.e., stable results are obtained for $\Lambda \rightarrow \infty$) and
the naively expected infinite number of counterterms
is not needed. This means that Weinberg power counting does actually work in
$S$ waves at LO (ignoring the $m_\pi$ dependence of the contact interaction
discussed in Refs.~\cite{KSW96,KSW98,KSW98a,Bea02}).
However, there are problems with a particular class of higher partial waves,
namely those  
in which the tensor force from 1PE is attractive. The first few cases
of this kind  of low angular momentum are
$^3P_0$, $^3P_2$, and $^3D_2$, which need a counterterm for cutoff independence. 
The leading order (nonderivative) counterterms do not contribute in
$P$ and higher waves, which is why Weinberg counting fails in these cases. 
But the second order contact potential provides counterterms
for $P$ waves. Therefore, the promotion
of, particularly, the $^3P_0$ and $^3P_2$ contacts from next-to-leading order (NLO) to LO would
fix the problem in $P$ waves. To take care of the $^3D_2$ problem,
a next-to-next-to-next-to-leading order (N$^3$LO) contact needs to be promoted to LO.
Partial waves with orbital angular momentum $L\geq 3$ may be calculated in Born
approximation with sufficient accuracy and, therefore, do not pose renormalization
problems.
In this way, one arrives at a scheme
of `modified Weinberg counting'~\cite{NTK05} for the leading order 
two-nucleon interaction.

For a quantitative chiral $NN$ potential one needs to advance all the way
to N$^3$LO~\cite{EM03}. Thus, the renormalization issue needs to be discussed beyond LO.
Naively, the most perfect renormalization procedure is the one where the cutoff
parameter $\Lambda$ is carried to infinity while stable results are maintained.
This was done successfully at LO in the work by Nogga {\it et al}~\cite{NTK05} described above.
At NNLO, the infinite-cutoff renormalization procedure has been investigated 
in~\cite{YEP07,YEP08,YEP09} for partial waves with total angular momentum $J\leq 1$ and
in~\cite{VA07} for all partial waves with $J\leq 5$. At N$^3$LO, only a study
of the $^1S_0$ state exists~\cite{Ent08}. Thus, a full analysis of the issue 
is still lacking.

It is, therefore, the purpose of this paper to study the method of (non-perturbative) infinite-cutoff renormalization 
systematically order-by-order from LO to N$^3$LO and for all partial-waves with $J\leq 6$.  
As discussed, it is necessary to carry this investigation through all of these orders, because the presently existing quantitative chiral $NN$ potentials are of order N$^3$LO and their renormalizability needs to be investigated.

This paper is organized as follows.
In Sec.~2, we will present the chiral $NN$ potential up to order N$^3$LO and, in Sec.~3, 
the non-perturbative renormalization will be discussed. The order by order convergence (or non-convergence)
is the subject of Sec.~4, and Sec.~5 will conclude the paper.

\section{The chiral $NN$ potential up to N$^3$LO}
\label{sec_NNpot}

EFTs are defined in terms of effective Langrangians which
are given by an infinite series of terms with increasing number of derivatives
and/or nucleon fields, with the dependence of each term on the pion field 
prescribed by the rules of broken chiral symmetry.
Applying this Lagrangian to a particular process, an unlimited number of Feynman 
graphs can be generated. Therefore,
we need a scheme that makes the theory manageable and calculabel.
This scheme
which tells us how to distinguish between large
(important) and small (unimportant) contributions
is ChPT, and
determining the power $\nu$ of the expansion
has become known as power counting.

Nuclear potentials are defined as sets of irreducible
graphs up to a given order.
The power $\nu$ of a few-nucleon diagram involving $A$ nucleons
is given in terms of naive dimensional analysis by:
\begin{equation} 
\nu = -2 +2A - 2C + 2L 
+ \sum_i \Delta_i \, ,  
\label{eq_nu} 
\end{equation}
with
\begin{equation}
\Delta_i  \equiv   d_i + \frac{n_i}{2} - 2  \, ,
\label{eq_Deltai}
\end{equation}
where $C$ denotes the number of separately connected pieces and
$L$ the number of loops in the diagram;
$d_i$ is the number of derivatives or pion-mass insertions and $n_i$ the number of nucleon fields 
(nucleon legs) involved in vertex $i$; the sum runs over all vertices contained
in the diagram under consideration.
Note that $\Delta_i \geq 0$
for all interactions allowed by chiral symmetry.
For an irreducible $NN$ diagram (``two-nucleon potential'', $A=2$, $C=1$),
Eq.~(\ref{eq_nu}) collapses to
\begin{equation} 
\nu =  2L + \sum_i \Delta_i \, .  
\label{eq_nunn} 
\end{equation}

Thus, in terms of naive dimensional analysis or ``Weinberg counting'',
the various orders of the irreducible graphs which define the chiral $NN$ potential 
are given by:
\beqa
V_{\rm LO} & = & 
V_{\rm ct}^{(0)} + 
V_{1\pi}^{(0)} 
\label{eq_VLO}
\\
V_{\rm NLO} & = & V_{\rm LO} +
V_{\rm ct}^{(2)} + 
V_{1\pi}^{(2)} +
V_{2\pi}^{(2)} 
\label{eq_VNLO}
\\
V_{\rm NNLO} & = & V_{\rm NLO} +
V_{1\pi}^{(3)} + 
V_{2\pi}^{(3)} 
\label{eq_VNNLO}
\\
V_{{\rm N}^3{\rm LO}} & = & V_{\rm NNLO} +
V_{\rm ct}^{(4)} +
V_{1\pi}^{(4)} +  
V_{2\pi}^{(4)} +
V_{3\pi}^{(4)} 
\label{eq_VN3LO}
\eeqa
where 
the superscript denotes the order $\nu$ of the low-momentum
expansion.
LO stands for leading order, NLO for next-to-leading
order, etc..
Contact potentials carry the subscript ``ct'' and
pion-exchange potentials can be identified by an
obvious subscript.

The charge-independent 1PE potential reads
\begin{equation}
V_{1\pi} ({\vec p}~', \vec p) = - 
\frac{g_A^2}{4f_\pi^2}
\: 
\boldtau_1 \cdot \boldtau_2 
\:
\frac{
\vec \sigma_1 \cdot \vec q \,\, \vec \sigma_2 \cdot \vec q}
{q^2 + m_\pi^2} 
\,,
\label{eq_1PEci}
\end{equation}
where ${\vec p}~'$ and $\vec p$ designate the final and initial
nucleon momenta in the center-of-mass system (CMS) and $\vec q \equiv
{\vec p}~' - \vec p$ is the momentum transfer; $\vec \sigma_{1,2}$ and
$\boldtau_{1,2}$ are the spin and isospin operators of nucleon 1 and
2; $g_A$, $f_\pi$, and $m_\pi$ denote axial-vector coupling constant,
the pion decay constant, and the pion mass, respectively. We use
$f_\pi=92.4$ MeV and $g_A=1.29$ throughout this work.  
Since higher order corrections contribute only to mass
and coupling constant renormalizations and since, on shell, there are
no relativistic corrections, the on-shell 1PE has the form
\eq{eq_1PEci} in all orders.

In this paper, we will specifically calculate neutron-proton ($np$) scattering 
and take the charge-dependence (isospin violation) of the 1PE into account.
Thus, the 1PE potential that we actually apply reads
\begin{equation}
V_{1\pi}^{(np)} ({\vec p}~', \vec p) 
= -V_{1\pi} (m_{\pi^0}) + (-1)^{T+1}\, 2\, V_{1\pi} (m_{\pi^\pm})
\,,
\label{eq_1penp}
\end{equation}
where $T$ denotes the isospin of the two-nucleon system and
\begin{equation}
V_{1\pi} (m_\pi) \equiv - \,
\frac{g_A^2}{4f_\pi^2} \,
\frac{
\vec \sigma_1 \cdot \vec q \,\, \vec \sigma_2 \cdot \vec q}
{q^2 + m_\pi^2} 
\,.
\end{equation}
We use $m_{\pi^0}=134.9766$ MeV and
 $m_{\pi^\pm}=139.5702$ MeV.

\subsection{Leading order (LO)}
The LO chiral $NN$ potential consists of a contact part and an 1PE part,
cf.\ Eq.~(\ref{eq_VLO}). The 1PE part is given by Eq.~(\ref{eq_1penp})
and the LO contacts are
\begin{equation}
V_{\rm ct}^{(0)}(\vec{p'},\vec{p}) =
C_S +
C_T \, \vec{\sigma}_1 \cdot \vec{\sigma}_2 \, ,
\label{eq_ct0}
\end{equation}
and, in terms of partial waves,
\beqa
V_{\rm ct}^{(0)}(^1 S_0)          &=&  \widetilde{C}_{^1 S_0} =
4\pi\, ( C_S - 3 \, C_T )
\nonumber \\
V_{\rm ct}^{(0)}(^3 S_1)          &=&  \widetilde{C}_{^3 S_1} =
4\pi\, ( C_S + C_T ) \,,
\label{eq_ct0_pw}
\eeqa
where $C_S, C_T, \widetilde{C}_{^1 S_0}, \widetilde{C}_{^3 S_1}$ are constants.

\subsection{Next-to-leading order (NLO)}
Multi-pion exchange starts at NLO and continues through
all higher orders. It involves
divergent loop integrals that need to be regularized.
An elegant way to do this is dimensional regularization
which 
(besides the main nonpolynomial result) 
typically generates polynomial terms with coefficients
that are, in part, infinite or scale dependent~\cite{KBW97}.
One purpose of the contacts is
to absorb all infinities and scale dependencies and make
sure that the final result is finite and scale independent.
This is the renormalization of the perturbatively calculated
$NN$ {\it potential}, which must be carefully distinguished from the non-perturbative
renormalization to be discussed in Sec.~\ref{sec_reno}.
The perturbative renormalization of the $NN$ potential
is very similar to what is done in the ChPT calculations
of $\pi\pi$ and $\pi N$ scattering, namely, a renormalization
order by order, which is the method of choice for any EFT.

For the NLO chiral $NN$ potential, Eq.~(\ref{eq_VNLO}), we need to specify
the second order contact part and the two-pion exchange (2PE) part.
The NLO contact terms are given by~\cite{ME11}
\beqa
V_{\rm ct}^{(2)}(\vec{p'},\vec{p}) &=&
C_1 \, q^2 +
C_2 \, k^2 
\nonumber 
\\ &+& 
\left(
C_3 \, q^2 +
C_4 \, k^2 
\right) \vec{\sigma}_1 \cdot \vec{\sigma}_2 
\nonumber 
\\
&+& C_5 \left( -i \vec{S} \cdot (\vec{q} \times \vec{k}) \right)
\nonumber 
\\ &+& 
 C_6 \, ( \vec{\sigma}_1 \cdot \vec{q} )\,( \vec{\sigma}_2 \cdot 
\vec{q} )
\nonumber 
\\ &+& 
 C_7 \, ( \vec{\sigma}_1 \cdot \vec{k} )\,( \vec{\sigma}_2 \cdot 
\vec{k} ) \,,
\label{eq_ct2}
\eeqa
with the partial-wave decomposition
\beqa
V_{\rm ct}^{(2)}(^1 S_0)          &=&  C_{^1 S_0} ( p^2 + {p'}^2 ) 
\nonumber \\
V_{\rm ct}^{(2)}(^3 P_0)          &=&  C_{^3 P_0} \, p p'
\nonumber \\
V_{\rm ct}^{(2)}(^1 P_1)          &=&  C_{^1 P_1} \, p p' 
\nonumber \\
V_{\rm ct}^{(2)}(^3 P_1)          &=&  C_{^3 P_1} \, p p' 
\nonumber \\
V_{\rm ct}^{(2)}(^3 S_1)          &=&  C_{^3 S_1} ( p^2 + {p'}^2 ) 
\nonumber \\
V_{\rm ct}^{(2)}(^3 S_1- ^3 D_1)  &=&  C_{^3 S_1- ^3 D_1}  p^2 
\nonumber \\
V_{\rm ct}^{(2)}(^3 D_1- ^3 S_1)  &=&  C_{^3 S_1- ^3 D_1}  {p'}^2 
\nonumber \\
V_{\rm ct}^{(2)}(^3 P_2)          &=&  C_{^3 P_2} \, p p'   \,.
\label{eq_ct2_pw}
\eeqa
To state the 2PE potentials, we introduce the following notation:
\begin{eqnarray} 
V_{2\pi}({\vec p}~', \vec p) &  = &
 \:\, V_C \:\, + 
\bbox{\tau}_1 \cdot \bbox{\tau}_2 
\, W_C 
\nonumber \\ &+&  
\left[ \, V_S \:\, + \bbox{\tau}_1 \cdot \bbox{\tau}_2 \, W_S 
\,\:\, \right] \,
\vec\sigma_1 \cdot \vec \sigma_2
\nonumber \\ &+& 
\left[ \, V_{LS} + \bbox{\tau}_1 \cdot \bbox{\tau}_2 \, W_{LS}    
\right] \,
\left(-i \vec S \cdot (\vec q \times \vec k) \,\right)
\nonumber \\ &+& 
\left[ \, V_T \:\,     + \bbox{\tau}_1 \cdot \bbox{\tau}_2 \, W_T 
\,\:\, \right] \,
\vec \sigma_1 \cdot \vec q \,\, \vec \sigma_2 \cdot \vec q  
\nonumber \\ &+& 
\left[ \, V_{\sigma L} + \bbox{\tau}_1 \cdot \bbox{\tau}_2 \, 
      W_{\sigma L} \, \right] \,
\vec\sigma_1\cdot(\vec q\times \vec k\,) \,\,
\vec \sigma_2 \cdot(\vec q\times \vec k\,)
\, ,
\label{eq_nnamp}
\end{eqnarray}
where 
\begin{equation}
\begin{array}{llll}
\vec k &\equiv& \frac12 ({\vec p}~' + \vec p) & \mbox{\rm is the 
average momentum, and}\\
\vec S &\equiv& \frac12 (\vec\sigma_1+\vec\sigma_2) & 
\mbox{\rm the total spin.}
\end{array}
\label{eq_defqk}
\end{equation}
Using the above notation, the NLO 2PE is simply given by~\cite{ME11,KBW97}
\begin{eqnarray} 
W_C^{(2)} &=&-{L(q)\over384\pi^2 f_\pi^4} 
\left[4m_\pi^2(5g_A^4-4g_A^2-1)
+q^2(23g_A^4 -10g_A^2-1) 
+ {48g_A^4 m_\pi^4 \over w^2} \right] \,,  
\label{eq_2C}
\\   
V_T^{(2)} &=& -{1\over q^2} V_{S}^{(2)} 
    \; = \; -{3g_A^4 L(q)\over 64\pi^2 f_\pi^4} \,, 
\label{eq_2T}
\end{eqnarray}  
where
\begin{equation} 
L(q)  \equiv  {w\over q} \ln {w+q \over 2m_\pi}
\end{equation}
and
\begin{equation} 
 w  \equiv  \sqrt{4m_\pi^2+q^2} \,. 
\end{equation}

\subsection{Next-to-next-to-leading order (NNLO)}
There are no new contacts at NNLO, cf.\ Eq.~(\ref{eq_VNNLO}), and, thus,
all we need is the third order 2PE potential, which is 
[using the notation introduced in Eq.~(\ref{eq_nnamp})]~\cite{ME11,KBW97}
\begin{eqnarray}
V_C^{(3)} &=& V_{C1}^{(3)} + V_{C2}^{(3)}\,,\\
W_C^{(3)} &=& W_{C1}^{(3)} + W_{C2}^{(3)}\,,\\
V_T^{(3)} &=& V_{T1}^{(3)} + V_{T2}^{(3)}\,,\\
W_T^{(3)} &=& W_{T1}^{(3)} + W_{T2}^{(3)}\,,\\
V_S^{(3)} &=& V_{S1}^{(3)} + V_{S2}^{(3)}\,,\\
W_S^{(3)} &=& W_{S1}^{(3)} + W_{S2}^{(3)}\,,\\
V_{LS}^{(3)} &=&  {3g_A^4  \widetilde{w}^2 A(q) \over 32\pi M_N f_\pi^4} 
 \,,\\  
W_{LS}^{(3)} &=& {g_A^2(1-g_A^2)\over 32\pi M_N f_\pi^4} 
w^2 A(q) \,, 
\end{eqnarray}
where
\begin{eqnarray} 
V_{C1}^{(3)} &=&{3g_A^2 \over 16\pi f_\pi^4} 
\left\{ {g_A^2 m_\pi^5  \over 16M_N w^2}  
-\left[2m_\pi^2( 2c_1-c_3)-q^2  \left(c_3 +{3g_A^2\over16M_N}\right)
\right]
\widetilde{w}^2 A(q) \right\} \,, 
\label{eq_3C}
\\
W_{C1}^{(3)} &=& {g_A^2\over128\pi M_N f_\pi^4} \left\{ 
 3g_A^2 m_\pi^5 w^{-2} 
 - \left[ 4m_\pi^2 +2q^2-g_A^2(4m_\pi^2+3q^2) \right] 
\widetilde{w}^2 A(q)
\right\} 
\,,\\ 
V_{T1}^{(3)} &=& -{1 \over q^2} V_{S1}^{(3)}
   \; = \; {9g_A^4 \widetilde{w}^2 A(q) \over 512\pi M_N f_\pi^4} 
 \,,  \\ 
W_{T1}^{(3)} &=&-{1\over q^2}W_{S1}^{(3)} 
    =-{g_A^2 A(q) \over 32\pi f_\pi^4}
\left[
\left( c_4 +{1\over 4M_N} \right) w^2
-{g_A^2 \over 8M_N} (10m_\pi^2+3q^2)  \right] 
\,,
\label{eq_3T}
\end{eqnarray}   
and
\begin{eqnarray}
V_{C2}^{(3)} &=& -\frac{3 g_A^4}{256 \pi f_\pi^4 M_N} 
(m_\pi w^2 + \widetilde w^4 A(q) )
\,,
\label{eq_3EM1}
\\
W_{C2}^{(3)} &=& \frac{g_A^4}{128 \pi f_\pi^4 M_N} 
(m_\pi w^2 + \widetilde w^4 A(q) )
\,,
\\
V_{T2}^{(3)} &=& -\frac{1}{q^2} V_{S2}^{(3)} = \frac{3 g_A^4}{512 \pi f_\pi^4 M_N} 
(m_\pi + w^2 A(q) )
\,,
\\
W_{T2}^{(3)} &=& -\frac{1}{q^2} W_{S2}^{(3)} = -\frac{g_A^4}{256 \pi f_\pi^4 M_N} 
(m_\pi + w^2 A(q) )
\,,
\label{eq_3EM4}
\end{eqnarray}
with
\begin{equation} 
A(q) \equiv {1\over 2q}\arctan{q \over 2m_\pi} 
\end{equation}
and
\begin{equation} 
\widetilde{w} \equiv  \sqrt{2m_\pi^2+q^2} \,. 
\end{equation}
Equations~(\ref{eq_3EM1})-(\ref{eq_3EM4}) are corrections of the iterative 2PE, see Ref.~\cite{ME11}
for details.
In all 2PE potentials, we use the average nucleon mass, $M_N=938.9182$ MeV, and the
average pion mass, $m_\pi=138.039$ MeV.
The values for the low-energy constants are
$c_1=-0.81$ GeV$^{-1}$,
$c_3=-3.20$ GeV$^{-1}$, and
$c_4=5.40$ GeV$^{-1}$~\cite{ME11}.

\subsection{Next-to-next-to-next-to-leading order (N$^3$LO)}
At N$^3$LO, 14 new contact terms appear~\cite{ME11}, 
\beqa
V_{\rm ct}^{(4)}(\vec{p'},\vec{p}) &=&
D_1 \, q^4 +
D_2 \, k^4 +
D_3 \, q^2 k^2 +
D_4 \, (\vec{q} \times \vec{k})^2 
\nonumber 
\\ &+& 
\left(
D_5 \, q^4 +
D_6 \, k^4 +
D_7 \, q^2 k^2 +
D_8 \, (\vec{q} \times \vec{k})^2 
\right) \vec{\sigma}_1 \cdot \vec{\sigma}_2 
\nonumber 
\\ &+& 
\left(
D_9 \, q^2 +
D_{10} \, k^2 
\right) \left( -i \vec{S} \cdot (\vec{q} \times \vec{k}) \right)
\nonumber 
\\ &+& 
\left(
D_{11} \, q^2 +
D_{12} \, k^2 
\right) ( \vec{\sigma}_1 \cdot \vec{q} )\,( \vec{\sigma}_2 
\cdot \vec{q})
\nonumber 
\\ &+& 
\left(
D_{13} \, q^2 +
D_{14} \, k^2 
\right) ( \vec{\sigma}_1 \cdot \vec{k} )\,( \vec{\sigma}_2 
\cdot \vec{k})
\nonumber 
\\ &+& 
D_{15} \left( 
\vec{\sigma}_1 \cdot (\vec{q} \times \vec{k}) \, \,
\vec{\sigma}_2 \cdot (\vec{q} \times \vec{k}) 
\right) \,,
\label{eq_ct4}
\eeqa
which contribute as follows to the partial-wave potentials,
\beqa
V_{\rm ct}^{(4)}(^1 S_0)          &=&  \widehat{D}_{^1 S_0}          
({p'}^4 + p^4) +
                              D_{^1 S_0}          {p'}^2 p^2 
\nonumber 
\\
V_{\rm ct}^{(4)}(^3 P_0)          &=&        D_{^3 P_0}          
({p'}^3 p + p' p^3) 
\nonumber 
\\
V_{\rm ct}^{(4)}(^1 P_1)          &=&        D_{^1 P_1}          
({p'}^3 p + p' p^3) 
\nonumber 
\\
V_{\rm ct}^{(4)}(^3 P_1)          &=&        D_{^3 P_1}          
({p'}^3 p + p' p^3) 
\nonumber 
\\
V_{\rm ct}^{(4)}(^3 S_1)          &=&  \widehat{D}_{^3 S_1}          
({p'}^4 + p^4) +
                              D_{^3 S_1}          {p'}^2 p^2 
\nonumber 
\\
V_{\rm ct}^{(4)}(^3 D_1)          &=&        D_{^3 D_1}          
{p'}^2 p^2 
\nonumber 
\\
V_{\rm ct}^{(4)}(^3 S_1 - ^3 D_1) &=&  \widehat{D}_{^3 S_1 - ^3 D_1} 
p^4             +
                              D_{^3 S_1 - ^3 D_1} {p'}^2 p^2
\nonumber 
\\
V_{\rm ct}^{(4)}(^3 D_1 - ^3 S_1) &=&  \widehat{D}_{^3 S_1 - ^3 D_1} 
{p'}^4             +
                              D_{^3 S_1 - ^3 D_1} {p'}^2 p^2
\nonumber 
\\
V_{\rm ct}^{(4)}(^1 D_2)          &=&        D_{^1 D_2}          
{p'}^2 p^2 
\nonumber 
\\
V_{\rm ct}^{(4)}(^3 D_2)          &=&        D_{^3 D_2}          
{p'}^2 p^2 
\nonumber 
\\
V_{\rm ct}^{(4)}(^3 P_2)          &=&        D_{^3 P_2}          
({p'}^3 p + p' p^3) 
\nonumber 
\\
V_{\rm ct}^{(4)}(^3 P_2 - ^3 F_2) &=&        D_{^3 P_2 - ^3 F_2} {p'}p^3
\nonumber 
\\
V_{\rm ct}^{(4)}(^3 F_2 - ^3 P_2) &=&        D_{^3 P_2 - ^3 F_2} {p'}^3p
\nonumber 
\\
V_{\rm ct}^{(4)}(^3 D_3)          &=&        D_{^3 D_3}          
{p'}^2 p^2   \,.
\label{eq_ct4_pw}
\eeqa
The 2PE contributions at this order, $V_{2\pi}^{(4)}$,  are very involved, which is why we will not reprint them here. 
The comprehensive expressions can be found in Appendix D of Ref.~\cite{ME11}. We note that, in the calculations
of this paper, we apply {\it all} N$^3$LO 2PE terms including those which require numerical integrations.
The parameters we use in this work are listed in column ``$NN$ potential'' of Table~2 of Ref.~\cite{ME11}.

The N$^3$LO three-pion exchange (3PE) contributions $V_{3\pi}^{(4)}$, cf.\ Eq.~(\ref{eq_VN3LO}),
are left out, since they have been found to be negligible~\cite{Kai00a,Kai00b}.

\section{$NN$ scattering and non-perturbative renormalization}
\label{sec_reno}

For the unitarizing scattering equation, we choose the relativistic three-dimensional equation
proposed by Blankenbecler and Sugar (BbS)~\cite{BS66},
which reads,
\begin{equation}
{T}({\vec p}~',{\vec p})= {V}({\vec p}~',{\vec p})+
\int \frac{d^3p''}{(2\pi)^3} \:
{V}({\vec p}~',{\vec p}~'') \:
\frac{M_N^2}{E_{p''}} \:  
\frac{1}
{{ p}^{2}-{p''}^{2}+i\epsilon} \:
{T}({\vec p}~'',{\vec p}) 
\label{eq_bbs2}
\end{equation}
with $E_{p''}\equiv \sqrt{M_N^2 + {p''}^2}$.
The advantage of using a relativistic scattering equation is that it automatically
includes relativistic corrections to all orders. Thus, in the scattering equation,
no propagator modifications are necessary when raising the order to which the
calculation is conducted.

Defining
\begin{equation}
\widehat{V}({\vec p}~',{\vec p})
\equiv 
\frac{1}{(2\pi)^3}
\sqrt{\frac{M_N}{E_{p'}}}\:  
{V}({\vec p}~',{\vec p})\:
 \sqrt{\frac{M_N}{E_{p}}}
\label{eq_minrel1}
\end{equation}
and
\begin{equation}
\widehat{T}({\vec p}~',{\vec p})
\equiv 
\frac{1}{(2\pi)^3}
\sqrt{\frac{M_N}{E_{p'}}}\:  
{T}({\vec p}~',{\vec p})\:
 \sqrt{\frac{M_N}{E_{p}}}
\,,
\label{eq_minrel2}
\end{equation}
where the factor $1/(2\pi)^3$ is added for convenience,
the BbS equation collapses into the usual, nonrelativistic
Lippmann-Schwinger (LS) equation, Eq.~(\ref{eq_LS}).
Since $\widehat V$ satisfies Eq.~(\ref{eq_LS}), 
it can be used like a usual nonrelativistic potential, and 
$\widehat{T}$ may be perceived as the conventional nonrelativistic T-matrix.
The square-root factors in Eqs.~(\ref{eq_minrel1}) and (\ref{eq_minrel2})
are applied to the potentials of all orders except in LO.

In the LS equation, \eq{eq_LS}, we use
\beqa
M_N  &=&  \frac{2M_pM_n}{M_p+M_n} = 938.9182 \mbox{ MeV, and}
\\
p^2 & = & \frac{M_p^2 T_{\rm lab} (T_{\rm lab} + 2M_n)}
               {(M_p + M_n)^2 + 2T_{\rm lab} M_p}  
\,,
\eeqa
where $M_p=938.2720$ MeV and $M_n=939.5653$ MeV
are the proton and neutron masses, respectively, and 
$T_{\rm lab}$ 
is the kinetic energy of the incident neutron 
in the laboratory system (``Lab.\ Energy'').
The relationship between $p^2$ and
$T_{\rm lab}$ 
is based upon relativistic kinematics.

We renormalize the LO chiral $NN$ potential as described in Refs.~\cite{NTK05,Mac09} and discussed in the
Introduction.
We then perform the infinite-cutoff renormalization also for the NLO, NNLO, and N$^3$LO chiral $NN$ potentials.
This is accomplished by studying the dependence of the phase shifts on
the cutoff parameter $\Lambda$ that appears in the regulator function, Eq.~(\ref{eq_regulator}).
We vary $\Lambda$ over a wide range, from $0.5\text{\ GeV}$ to $10\text{\ GeV}$ 
(with $n=2$ for the LO, NLO, and NNLO potentials, and $n=3$ for N$^3$LO).

For partial-waves with short-range repulsion, convergence with increasing cutoff values is obtained without the use of any counterterm, i.e., convergence occurs \lq\lq automatically".  In fact, in these cases, any counterterm becomes ineffective for large cutoffs.  Thus, no counterterm is used in partial-waves with short-range repulsion.

For partial-waves with short-range attraction, one counterterm (contact term) is needed to ensure convergence.  If we introduce a second counterterm per partial-wave, it turns out that this second parameter becomes ineffective for large cutoffs.  Thus, we apply only one counterterm, which we use to fit the following empirical information.  In $S$-waves, we fit the scattering lengths; $^{1}S_{0}: a_{s}=-23.740\text{ fm}$, $^{3}S_{1}: a_{t}=5.417\text{ fm}$.  In the other partial-waves with $J\leq2$, we fit the phase shift at $T_{\rm lab}=50\text{ MeV}$ to the central value from the Nijmegen multi-energy $np$ phase-shift analysis~\cite{Sto93}.  For $J\geq3$, we fit the phase-shift at $100\text{ MeV}$ or $200\text{ MeV}$.

In summary, at any order, either one or no counterterm is needed in each partial-wave for the infinite-cutoff renormalization.  We show this in the left half of Table~\ref{tbl_pw_ct} for the various partial-wave states.  The right half of Table~\ref{tbl_pw_ct} shows the number of counterterms according to Weinberg Counting.  Obviously, there are large differences between the two schemes.  The higher partial-wave states that are not listed in the table do not receive counterterms.

\begin{table}[t]
\vspace*{0cm}
\begin{center}
\caption{Number of counterterms per partial-wave as required in two different renormalization schemes.}
\begin{tabular}{|c||c|c|c|c||c|c|c|}
\hline
 & \multicolumn{4}{|c||}{Infinite-cutoff renormalization}
& \multicolumn{3}{|c|}{Weinberg Counting}\\
\hline
Partial-Wave & LO & NLO & NNLO & N$^3$LO & LO & NLO/NNLO & N$^3$LO \\
\hline\hline
$^{1}S_{0}$ & 1 & 1 & 1 & 1 & 1 & 2 & 4\\
$^{3}P_{0}$ & 1 & 0 & 0 & 0 & 0 & 1 & 2\\
\hline
$^{1}P_{1}$ & 0 & 0 & 0 & 0 & 0 & 1 & 2\\
$^{3}P_{1}$ & 0 & 1 & 1 & 1 & 0 & 1 & 2\\
$^{3}S_{1}$ & 1 & 0 & 1 & 1 & 1 & 2 & 4\\
$^{3}S_{1}- ^{3}D_{1}$ & 0 & 0 & 1 & 0 & 0 & 1 & 3\\
$^{3}D_{1}$ & 0 & 0 & 0 & 0 & 0 & 0 & 1\\
\hline
$^{1}D_{2}$ & 0 & 1 & 1 & 1 & 0 & 0 & 1\\
$^{3}D_{2}$ & 1 & 0 & 1 & 0 & 0 & 0 & 1\\
$^{3}P_{2}$ & 1 & 1 & 1 & 1 & 0 & 1 & 2\\
$^{3}P_{2}- ^{3}F_{2}$ & 0 & 0 & 0 & 0 & 0 & 0 & 1\\
$^{3}F_{2}$ & 0 & 0 & 0 & 0 & 0 & 0 & 0\\
\hline
$^{1}F_{3}$ & 0 & 0 & 0 & 0 & 0 & 0 & 0\\
$^{3}F_{3}$ & 0 & 0 & 1 & 1 & 0 & 0 & 0\\
$^{3}D_{3}$ & 0 & 0 & 1 & 0 & 0 & 0 & 1\\
$^{3}D_{3}- ^{3}G_{3}$ & 0 & 0 & 0 & 0 & 0 & 0 & 0\\
$^{3}G_{3}$ & 0 & 0 & 0 & 0 & 0 & 0 & 0\\
\hline
$^{1}G_{4}$ & 0 & 0 & 1 & 1 & 0 & 0 & 0\\
$^{3}G_{4}$ & 0 & 0 & 0 & 0 & 0 & 0 & 0\\
$^{3}F_{4}$ & 0 & 0 & 1 & 1 & 0 & 0 & 0\\
\hline
\end{tabular}
\label{tbl_pw_ct}
\end{center}
\vspace*{0cm}
\end{table}
\pagebreak

\begin{figure}
\vspace*{-2.0cm}
\hspace{-1.0cm}
\includegraphics[scale=.25]{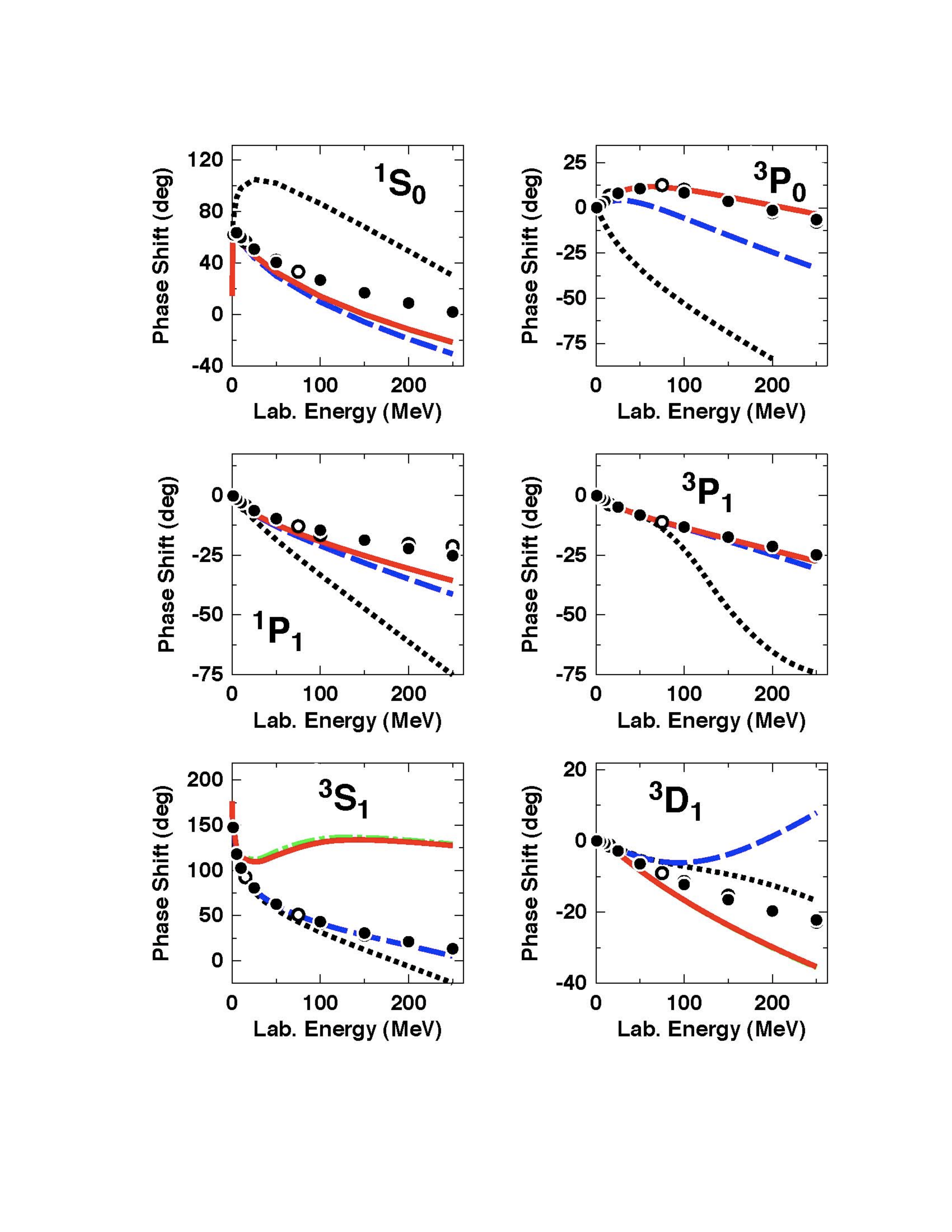}
\vspace*{-2.5cm}
\caption{Phase-shifts of neutron-proton scattering at order N$^3$LO for total angular momentum $J\leq1$ and laboratory kinetic energies below $250\text{\ MeV}$.  The black dotted curve is obtained for $\Lambda=0.5\text{\ GeV}$, the blue dashed curve for $\Lambda=1\text{\ GeV}$, the green dash-dot curve for $\Lambda=5\text{\ GeV}$, and the red solid curve for $\Lambda=10\text{\ GeV}$.  Note that the curves for $\Lambda=5\text{\ GeV}$ and $10\text{\ GeV}$ are, in general, indistinguishable on the scale of the figure.  The filled and open circles represent the results from the Nijmegan multi-energy $np$ phase-shift analysis~\cite{Sto93} and the VPI/GWU single-energy $np$ analysis SM99~\cite{SM99}, respectively.}
\label{fig_51}
\end{figure}

\begin{figure}
\vspace*{-2.0cm}
\hspace{-1.0cm}
\includegraphics[scale=.25]{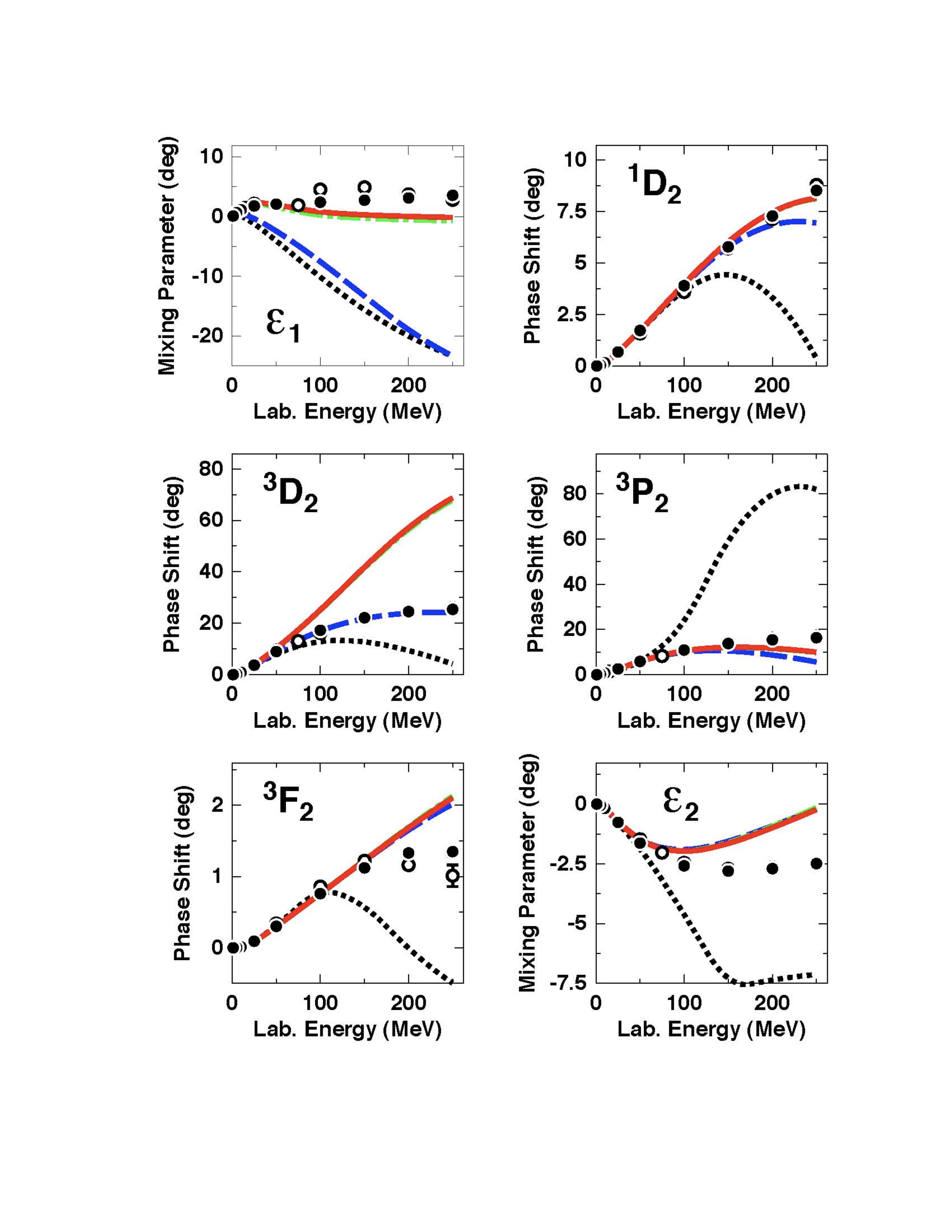}
\vspace*{-2.5cm}
\caption{Same as Fig.~\ref{fig_51}, but $J=2$ phase-shifts and $J\leq2$ mixing parameters are shown}
\label{fig_52}
\end{figure}
\pagebreak

We have obtained convergence with increasing cutoff for all phase parameters with $J\leq 6$ in each order, NLO, NNLO, and N$^3$LO.  As evident from Table~\ref{tbl_pw_ct}, in NNLO and N$^3$LO, we need counterterms 
(contact terms) for the renormalization of the  $^3F_3$, $^3F_4$, and $^1G_4$ waves.  These terms are:
\begin{eqnarray}
V_{ct}^{(6)}(^{3}F_{3}) & = & E_{^{3}F_{3}}p'^{3}p^{3}, \\
V_{ct}^{(6)}(^{3}F_{4}) & = & E_{^{3}F_{4}}p'^{3}p^{3}, \\
V_{ct}^{(8)}(^{1}G_{4}) & = & F_{^{1}G_{4}}p'^{4}p^{4}.
\end{eqnarray}

In Figures~\ref{fig_51} and~\ref{fig_52}, we demonstrate the convergence of the phase parameters with increasing cutoff for the N$^3$LO potential in partial-wave with $J\leq2$.  The black dotted curve is obtained for $\Lambda=0.5\text{\ GeV}$, the blue dashed curve for $\Lambda=1\text{\ GeV}$, the green dash-dot curve for $\Lambda=5\text{\ GeV}$, and the red solid curve for $\Lambda=10\text{\ GeV}$.  The fact that the green and red curves are essentially indistinguishable in those figures, demonstrates that convergence for large cutoffs has occured.
However, it is also clearly seen in these figures that, even though we are here at a relatively high order (N$^3$LO), 
the cutoff-converged curves
show large discrepancies with respect to the empirical phase shifts in several partial waves, particularly,
$^1S_0$, $^3S_1$, $^3D_1$, $^3D_2$, and $\epsilon_2$. This issue will be further discussed in the
next section.

\section{Order By Order Convergence}
\label{sec_conv}

\begin{figure}
\vspace*{-2.0cm}
\hspace*{-1.0cm}
\includegraphics[scale=.8]{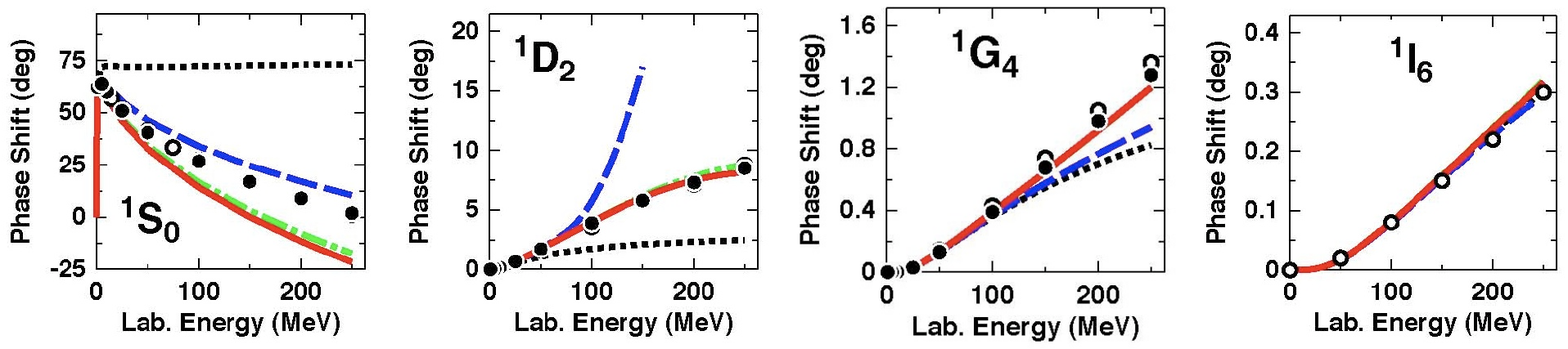}
\vspace*{-17cm}
\caption{Renormalized $np$ phase-shifts at order LO (black dotted curve), NLO (blue dashed curve), NNLO (green dash-dotted curve), and N$^3$LO (red solid curve).  The $S=0$, $T=1$ phase shifts with $L\leq6$ are shown for energies below $250\text{\ MeV}$.  Filled and open circles are as described in Fig.~\ref{fig_51}.}
\label{fig_53}
\end{figure}

\begin{figure}
\vspace*{-1.3cm}
\hspace*{-0.7cm}
\includegraphics[scale=.7]{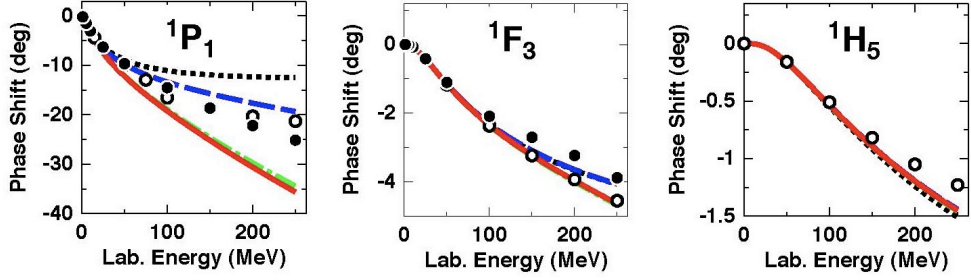}
\vspace*{-13.5cm}
\caption{Renormalized $S=0$, $T=0$ $np$ phase-shifts with $L\leq5$.  Notation as in Fig.~\ref{fig_53}.}
\label{fig_54}
\end{figure}

\begin{figure}
\vspace*{-1.3cm}
\hspace*{-0.7cm}
\includegraphics[scale=.7]{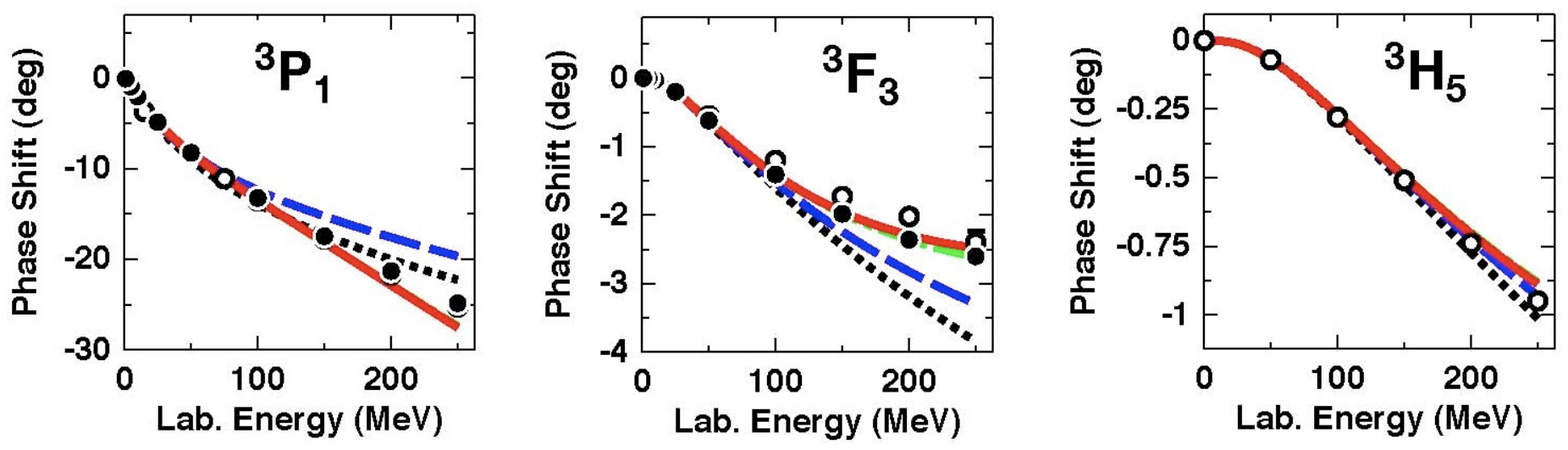}
\vspace*{-13.5cm}
\caption{Renormalized $S=1$, $T=1$ uncoupled $np$ phase-shifts with $L\leq5$.  Notation as in Fig.~\ref{fig_53}.}
\label{fig_55}
\end{figure}

\begin{figure}
\vspace*{-1.4cm}
\hspace*{-0.7cm}
\includegraphics[scale=.7]{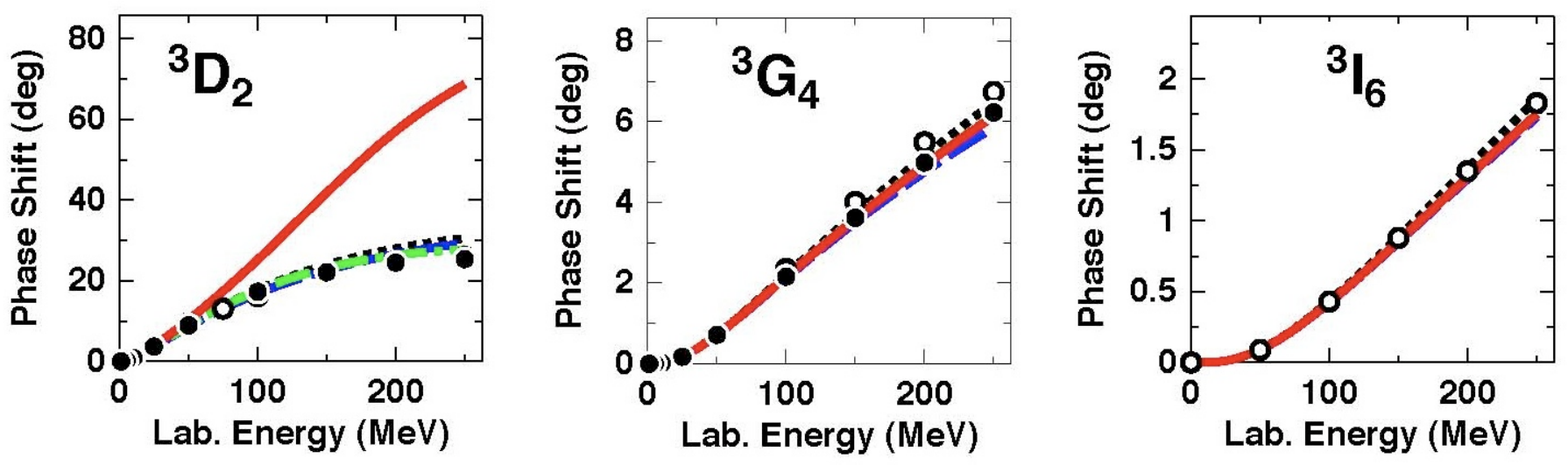}
\vspace*{-13.5cm}
\caption{Renormalized $S=1$, $T=0$ uncoupled $np$ phase-shifts for $L\leq6$.  Notation as in Fig.~\ref{fig_53}.}
\label{fig_56}
\end{figure}

\begin{figure}
\vspace*{-2.0cm}
\hspace*{-0.7cm}
\includegraphics[scale=.7]{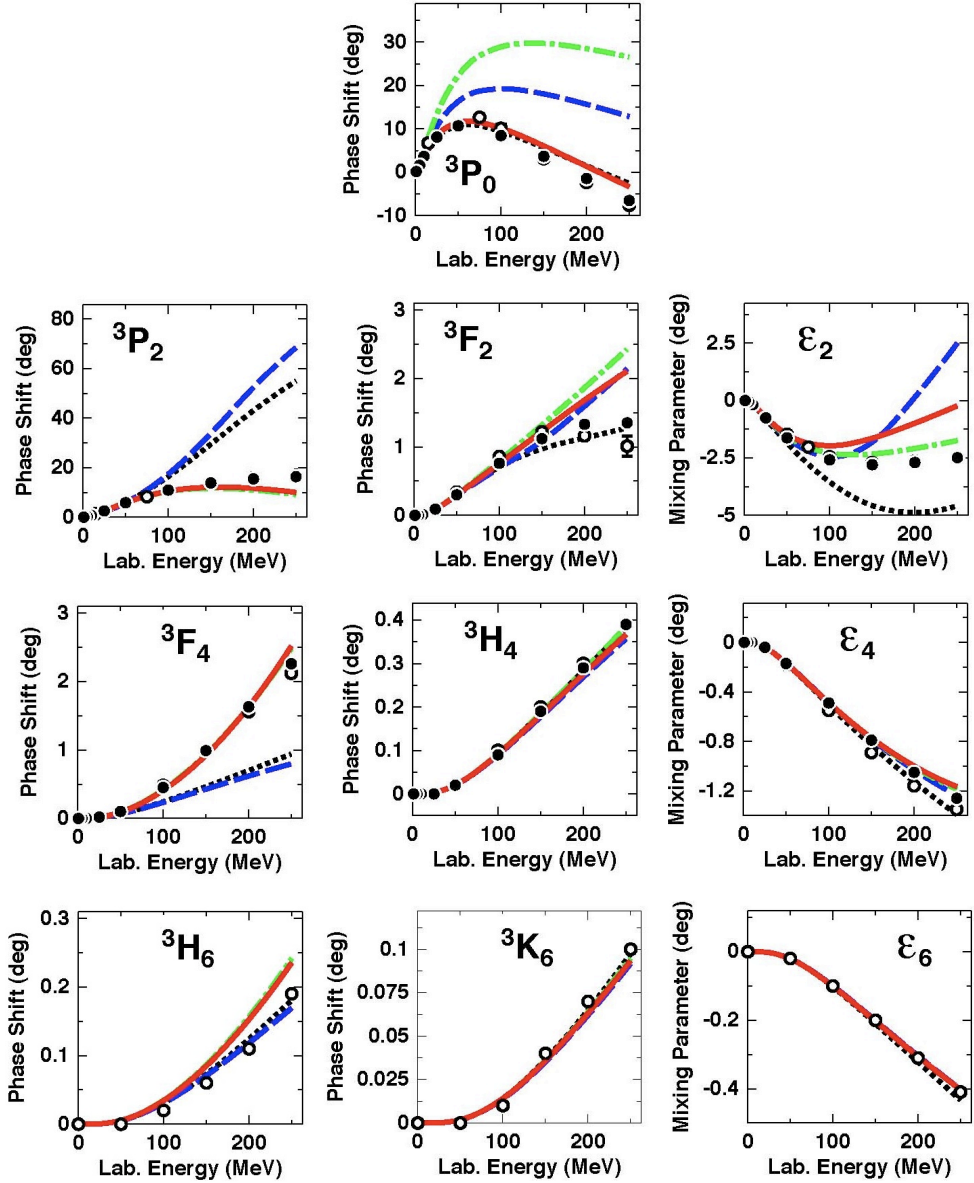}
\vspace*{-1.5cm}
\caption{The renormalized $^3P_{0}$ and renormalized $S=1$, $T=1$ coupled $np$ phase parameters for $J\leq6$.  Notation as in Fig.~\ref{fig_53}.}
\label{fig_57}
\end{figure}

\begin{figure}
\vspace*{-2.0cm}
\hspace*{-0.7cm}
\includegraphics[scale=.7]{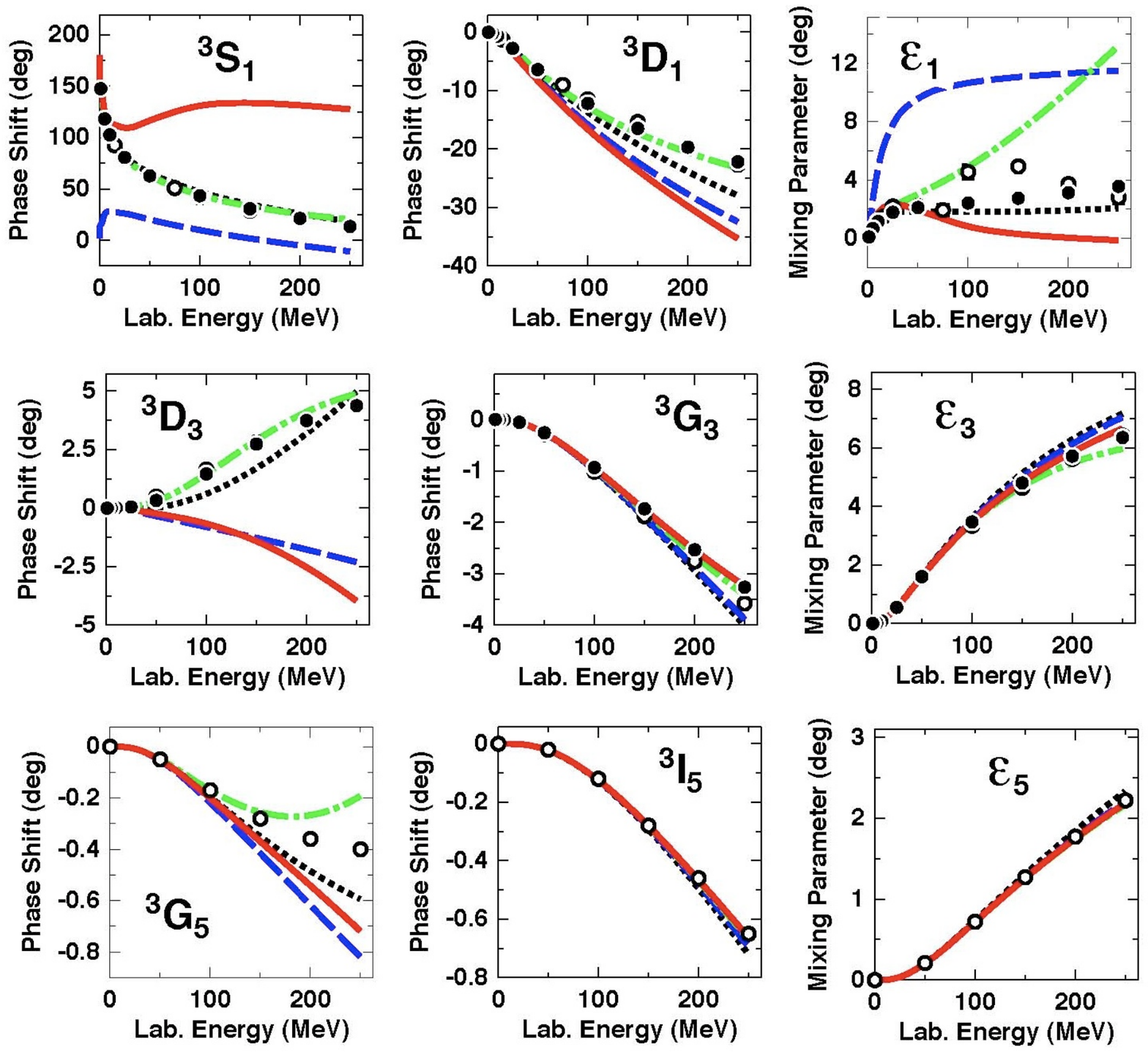}
\vspace*{-5.5cm}
\caption{Renormalized $S=1$, $T=0$ coupled $np$ phase parameters for $J\leq5$.  Notation as in Fig.~\ref{fig_53}.}
\label{fig_58}
\end{figure}

Having accomplished the infinite-cutoff renormalization of the four orders we consider, it is now of interest to investigate the order-by-order convergence of these cutoff-converged cases.  This is done in Figs.~\ref{fig_53}-\ref{fig_58}.

\pagebreak

In Fig.~\ref{fig_53}, the $S=0$, $T=1$ $np$ phase-shifts for $L\leq6$ are displayed, where $S$ denotes the total spin, $T$ the total isospin, and $L$ the orbital angular momentum of the two-nucleon system.  Note that, in each order, the underlying analytic expression for the potential is the same except that it is decomposed into partial-waves with $L=0$, $2$, $4$, and $6$.  The effect of this partial-wave decomposition is that the short-range part of the two-nucleon potential is increasingly suppressed with growing $L$; or in other words, the \lq\lq centrifugal barrier" becomes larger with $L$.  In the $^{1}S_{0}$ state, the phase-shifts are (almost) converged at N$^3$LO, but do not reproduce the empirical phase-shifts.  This is consistent with what was found in Ref.~\cite{Ent08}.  However, at the next higher $S=0$, $T=1$ partial-wave, the $^1D_{2}$, we observe that the phase-shift prediction converges to the empirical phase-shift values, and this is also true for all higher $S=0$, $T=1$ partial-waves shown in Fig.~{\ref{fig_53}}.

In Fig.~{\ref{fig_54}}, we show the $S=0$, $T=0$ state for various partial-waves up to $L=5$.  The situation is very similar to what we just discussed.  In the lowest partial-wave, the $^1P_{1}$, we have (almost) convergence, though not to the empirical values.  In higher partial-waves, however, the predictions converge to the experimental phase-shifts.  This finishes the discussion of (spin) singlet states ($S=0$).

Now we turn to (spin) triplet states ($S=1$).  Here, we need to distinguish between uncoupled and coupled partial-waves and will discuss the former first.  The phase-shift predictions for uncoupled $S=1$, $T=1$ states are shown in Fig.~{\ref{fig_55}} and for $S=1$, $T=0$ in Fig.~{\ref{fig_56}}.  Good convergence to the empirical phase-shifts is observed for all $S=1$, $T=1$ partial-waves (including the lowest one, $^3P_{1}$).  In contrast, for $S=1$, $T=0$, a strong divergence occurs at N$^3$LO in the $^3D_{2}$ state (first frame of Fig.~{\ref{fig_56}}), while the higher $S=1$, $T=0$ partial-waves converge well.

Finally, we discuss the coupled cases.  The $T=1$ coupled partial-waves are displayed in Fig.~{\ref{fig_57}}.  We include here the $^3P_{0}$ state, such that we can compare it with its counterparts of larger $L$, namely $^3F_{2}$, $^3H_{4}$, and $^3K_{6}$.  It is clearly seen that the $^3P_{0}$ shows no order-by-order convergence, while the $^3F_{2}$ has near convergence and the $^3H_{4}$ and $^3K_{6}$ states are fully converged to the empirical information.  The associated coupled partial-waves show corresponding trends.

The last set of partial-waves to be discussed are the $S=1$, $T=0$ coupled states shown in Fig.~{\ref{fig_58}}.  While there is no convergence with increasing orders for the $J=1$ states, converged results and agreement with the experimental parameters are seen in the $L=J+1$ partial-waves ($^3G_{3}$ and $^3I_{5}$) and in the mixing parameters with $J\geq3$ (i.e., $\varepsilon_{3}$ and $\varepsilon_{5}$).  Contrary to this, the $L=J-1$ waves ($^3S_{1}$, $^3D_{3}$, $^3G_{5}$) never converge.  Even in the rather high partial-wave, $^3G_{5}$, there is a large difference between the NNLO and N$^3$LO predictions.  This phenomenon may be related to the fact that the $^3G_{5}$ is the only higher partial-wave that disagrees with the empirical phase-shifts at N$^3$LO, as was noticed already in Ref.~\cite{EM02a}.  At the present time, we do not have an explanation for this problem.

The observations we have made in conjunction with Figs.~{\ref{fig_53}-\ref{fig_58}} can be summarized as follows:  Some lower partial-waves show no convergence and no order-by-order improvement.  On the other hand, all higher partial-waves (except for the notorious $^3G_{5}$)  do not only converge, they also converge to the empirical values.  Note that in higher partial-waves the short-range part of the $NN$ interaction (equivalent to the high-momentum components of the interaction) is suppressed by the centrifugal barrier.  This suggests that the long- and intermediate-range part of the $NN$ potential up to N$^3$LO is reasonable, while the short-range part may be, in part, un-physical.

In the infinite-cutoff renormalization the potential is admitted up to unlimited momenta.  However, the EFT this potential is derived from has validity only for momenta smaller than the chiral symmetry breaking scale $\Lambda_{\chi}\approx1\text{\ GeV}$.  The lack of order-by-order convergence and the discrepancies in lower partial-waves demonstrate that the potential should not be used beyond the limits of the effective theory (see Ref.~\cite{EG09}
for a related discussion).  The conclusion then is that cutoffs should be limited to $\Lambda\lesssim\Lambda_{\chi}$.

\section{Summary and Conclusions}
\label{sec_concl}

We have investigated a particular scheme for the non-perturbative renormalization of the nucleon-nucleon ($NN$) potential based upon chiral effective field theory (chiral EFT).  In the scheme applied, the cutoff parameter of the regulator is taken to infinity.

Two vital requirements for constituting the legitimacy of an EFT are regulator independence and a power counting scheme that allows for order-by-order improvements of the predictions.  We were able to achieve regulator independence (i.e., cutoff-independence) for all orders of chiral perturbartion theory  considered (i.e., 
LO, NLO, NNLO, and N$^3$LO) and all partial-waves up to $J=6$.  In general, cutoff-independence is seen for cutoff values above $5{\text{\ GeV}}$.

However, in this investigation, we have also observed that large cutoffs impose limitations on the effectiveness of counterterms.  In each partial-wave, either no counterterm (case of short-range repulsion) or one counterterm (case of short-range attraction) is effective.  Therefore, the power counting scheme implied by the infinite-cutoff method is considerably different from the one of Naive Dimensional Analysis  or Weinberg Counting (cf.~Table~\ref{tbl_pw_ct}).  As a consequence, order-by-order improvements of the predictions do not occur in several lower partial-waves.

Thus, the chiral EFT approach to nuclear forces fails when renormalized by the infinite-cutoff method.  This result may not come as a surprise considering that the EFT which we apply is designed for momenta below the chiral-symmetry breaking scale, $\Lambda_{\chi}\approx1{\text{\ GeV}}$.  Under the infinite-cutoff method, the potential contributes for momenta that are far beyond the hard scale of $1{\text{\ GeV}}$. Our results suggest that finite-cutoffs $\lesssim1{\text{\ GeV}}$ should be used for the non-perturbative regularization of the chiral $NN$ potential.  With such cutoffs, all
counterterms of Weinberg Counting are effective, and an order-by-order improvement of the predictions is to be
expected~\cite{Lep97}.

\begin{acknowledgements}
The authors gratefully acknowledge enlightening discussions with E. Ruiz Arriola and M. Pavon Valderrama.
The work by R. M. was supported in part by the U.S. Department of Energy
under Grant No.~DE-FG02-03ER41270.
The work of D. R. E. was funded by the Ministerio de Ciencia y
Tecnolog\'\i a under Contract No.~FPA2007-65748, the Junta de Castilla
y Le\'on under Contract No.~GR12,  and
the European Community-Research Infrastructure Integrating
Activity ``Study of Strongly Interacting Matter'' (HadronPhysics2
Grant No.~227431).
\end{acknowledgements}



\end{document}